\documentstyle[psfig]{mn2e}
\input{psfig.sty}

\newif\ifAMStwofonts

\def\Mesz{M\'esz\'aros~}
\def\Pacz{Paczy\'nski~}

\def\p{$e^\pm \;$}
\def\msun{M$_{\odot}$}
\begin{document}

\title[Jets, winds  and bursts from coalescing
       neutron stars]  
        {Jets, winds and bursts from coalescing
        neutron stars}    
\author[Rosswog \& Ramirez-Ruiz]
        {Stephan Rosswog$^{1}$ and Enrico Ramirez-Ruiz$^{2}$
\\${\bf 1.}$ Department of Physics and Astronomy, University of
Leicester, LE1 7RH, Leicester, UK.
\\${\bf 2.}$ Institute of Astronomy, Madingley Road, Cambridge, CB3
        0HA, UK.}

\date{}

\maketitle

\label{firstpage}

\begin{abstract}
Recent high-resolution calculations of neutron star coalescences
are used to investigate whether $\nu \overline{\nu}$ 
annihilation can provide sufficient energy to   power gamma-ray
bursts, especially those belonging to the short duration category. 
Late time slices of the simulations, where the neutrino emission
has reached a maximum, stationary level are analyzed to address
this question.
We find  that $\nu \overline{\nu}$  annihilation can provide the
necessary driving stresses that lead to relativistic jet
expansion. Maximum Lorentz factors of the order of 15 are found
along the binary rotation axis, but larger values are expected 
to arise from higher numerical resolution. Yet the accompanying neutrino-driven
wind must be absent from the axis when the burst occurs or prohibitive
baryon loading may occur. We argue that even under the
most favorable conditions,  $\nu \overline{\nu}$
annihilation is unlikely to power a burst in excess of  $\sim
10^{48}$ erg. Unless the emission is beamed into less than
one percent of the solid angle, which we argue is improbable if it
is collimated by gas pressure,  this may fail to satisfy the apparent
isotropic energies inferred at cosmological distances.  
This mechanism may nonetheless  drive
a precursor fireball, thereby evacuating a cavity into which a later
magnetic driven jet could expand. A large range of time
delays between the merger 
and the black hole formation are to be expected. 
If the magnetic driven jet occurs after the black hole has formed,
a time span as long as weeks could pass between the neutrino powered precursor
and the magnetic driven GRB.         
\end{abstract}


\section{Introduction}

Double neutron star (NS) binaries, such as the famous  PSR1913+16,
will eventually 
coalesce due to angular momentum and energy losses to gravitational 
radiation, provided that their orbital separation is small enough. When a NS 
binary coalesces,
the rapidly-spinning merged system could be too massive to form a
single NS; on the other hand, the total angular momentum is probably
too large to be swallowed immediately by a black hole (BH). The expected
outcome would then be a spinning BH, orbited by a torus of 
NS debris,  whose accretion can 
release sufficient gravitational energy, $\approx 10^{52}$ erg, to
power a gamma-ray burst (GRB; Lattimer \& Schramm 1976;
Paczy\'{n}ski 1986, 1991; Eichler at al. 1989; Narayan, Paczy\'{n}ski
\& Piran 1992; Mochkovitch et al. 1993; Klu\'{z}niak \& Lee 1998; Rees 1999;
Rosswog et al. 1999; Ruffert \& Janka 1999; Salmonson,
Wilson \&  Mathews 2001; Rosswog \& Davies 2002). 
The energy released in the accretion process is expected to be transformed
into a fireball,
which may provide the driving stress necessary for relativistic
expansion (see Piran 1999 and \Mesz 2002 for recent reviews). 
Possible forms of this
outflow include the following. Relativistic particles generated by
$\nu\overline{\nu}$ annihilation -- the thermal energy released as 
neutrinos is reconverted, via collisions, into $e^{\pm}\;$ pairs and
photons. Electromagnetic Poynting flux -- as in pulsars, strong magnetic fields
anchored in the dense matter convert the rotational energy of the
system into a Poynting-dominated outflow. In either
case, the duration of the burst is determined by the 
viscous timescale of the accreting gas, which is substantially longer
than the dynamical or orbital timescale, thus providing a  simple
interpretation for the large difference between
the durations of bursts and their fast variability.\\
These binary encounters are thought to
transpire outside star forming regions (Fryer, Woosley \& Hartmann
1999) and are not currently thought to be appropriate for long
GRBs 
($\ge 20$ s), whose
afterglows have been predominantly localised within the optical image
of the host galaxy (Bloom, Kulkarni \& Djorgovski 2001). However, these
calculations are uncertain because they are sensitive to a number of
parameters that are not well understood (e.g. distribution of initial
separations; see Belczynski et al. 2002). Such models may
also be in difficulties producing adequate  duration, collimation, and
energy (Katz \& Canel 1996; Ruffert et 
al. 1997; Popham, Woosley \& Fryer 1999; Ruffert \& Janka 1999; Lee \&
Ramirez-Ruiz 2002), but each of these problems is model-dependent and
might be cured  by different assumptions regarding the poorly known
input physics such as disk viscosity or the jet production mechanism.\\   
In this Letter, we study  the neutrino emission produced in the shocked and 
dissipatively heated merger remnant of a NS binary encounter. 
Late time slices of three dimensional (3D), high resolution calculations 
of the last stages of the coalescence are used (Rosswog \& Davies 2002; 
hereafter RD) to investigate
the potential of $\nu\overline{\nu}$ annihilation as a viable source
for GRB production, at least for bursts  belonging to the short
category. Neutrinos could give rise to a relativistic $e^{\pm}\;$
pair dominated outflow if they annihilate in
a region of low enough baryon density. To that effect, high
resolution simulations seem to be
required, since even a very small amount of baryons polluting the
outflow could severely limit the attainable Lorentz
factor. Information regarding how much
$\nu\overline{\nu}$ energy and momentum has been injected, its
confinement and collimation  
are at the forefront of our attention. We address all three 
issues here, along with the type of predictions that would help us to 
discriminate between this mode of energy extraction and that of strong
magnetic fields anchored in the dense matter.
\section{Numerical Methods}
 
The 3D hydrodynamics simulations of the mergers of
binary NS were performed by RD using
a smoothed particle hydrodynamics method with up to $\sim 10^6$
particles. A realistic equation of state for hot, dense nuclear matter
has been used, which is based on the tables provided
by (Shen et al. 1998a,b) and smoothly extended to the  
low density regime with a gas consisting of neutrons, alpha particles, photons
and $e^{\pm}\;$ pairs (see RD for details).  
This allows one to closely follow the
thermodynamic evolution of the neutron-rich debris after the
merger. Under the conditions encountered  in the debris --
temperatures of several  
MeV and densities around $\sim 10^{12}$ g cm$^{-3}$ -- neutrinos are
emitted  copiously.  Their effect on the cooling and the changes in
the composition   of the material are taken into 
account via a detailed multi-flavour neutrino treatment which is described in 
detail in Rosswog et al. (2002). The results presented here are based
on the close analysis of late time segments, 
typically a time $t_{\rm sim} \approx 15$ ms after the start of the
merger simulations, where the neutrino luminosities
have reached their maximum, stationary level (Rosswog et al. 2002).
Since we are interested in the maximum possible effect, we assume
the system to maintain this level of emission for as long as 1 s (see \S 3). 
We describe results from three representative runs. First, an
initially corotating system with twice 1.4 M$_{\odot}$ (run D from
RD, referred to in the following as c1.4),  
which yields the lowest temperatures of all runs; it can therefore be 
considered as a lower limit on the neutrino luminosities. Second, a system 
with twice 1.4 M$_{\odot}$ and no initial NS spins (``irrotational'';
run C from RD, hereafter i1.4), which we  
regard as the generic case and finally, an extreme system of twice 2.0 
M$_{\odot}$ and no initial spins (run E from RD, hereafter i2.0),
which we consider  to be an upper limit for the neutrino luminosity.\\ 

\section{\p Fireballs produced by $\nu\overline{\nu}$ annihilation}

The $\nu\overline{\nu} \rightarrow e^{+}e^{-}$ process (which scales
with the square of the neutrino luminosity) can tap the
thermal energy of the hot debris disk.  
For this mechanism to be efficient, the neutrinos must  both escape
before being advected and not being produced too gradually. Typical
estimates for coalescing NS suggest a limit of $\le 10^{51}$ erg 
for the neutrino energy dumped into \p-pairs (Ruffert et al. 1997; 
Ruffert \& Janka 1999; Popham et al. 1999). 
If the \p-dominated plasma were 
collimated into a solid angle $\Omega_{\nu\overline{\nu}}$, then of
course the apparent  ``isotropized'' energy would be larger
by a factor of $4\pi/\Omega_{\nu\overline{\nu}}$.

The presence of a
region of very low density along the rotation axis, away from
the equatorial plane of the debris must be present  when the burst
occurs or otherwise prohibitive baryon loading will occur -- for instance  an
energy of $10^{53}$ erg deposited into \p pairs could not accelerate an
outflow to $\Gamma > 100$ if it had to drag more than $E/(\Gamma c^2)
\approx 5 \cdot 10^{-4} M_\odot$ baryons with it.
\begin{figure}
\vspace*{-2.5cm}
\centerline{\psfig{figure= 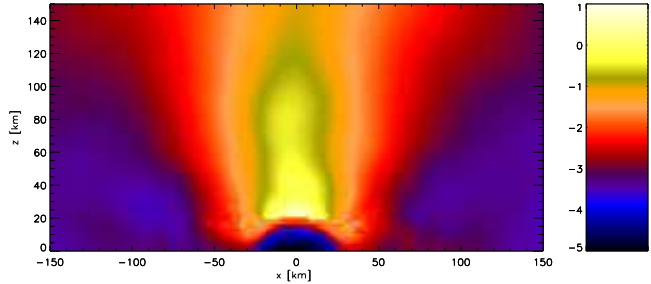,angle=90,width=9.cm}}
{\caption{Colour-coded is the ratio of energy deposited via 
$\nu\overline{\nu} \rightarrow e^{+}e^{-}$ to rest mass energy, $\eta$,
which is a measure of the maximum attainable Lorentz factor.
Shown are the values of log($\eta$) in the x-z-plane above the merged 
remnant of model i1.4 (no initial spins, 2 x 1.4 \msun) at the end of 
the simulation (t=13.8 ms). Due to the symmetry of the merger remnant
with respect to the orbital plane a similar jet will occur along the 
negative z-axis.}
\label{fig1}}
\end{figure}
In Figure 1 we plot the ratio of \p energy deposition
to baryon rest mass energy, $\eta=Q_{\nu\overline{\nu}}\tau_{\rm inj}/(\rho
c^2)$, in the region above the poles of the merged remnant. 
$\rho$ denotes the matter density and, for simplicity, an
energy injection  time, $\tau_{\rm inj}= 1$ s, well within the
distribution of the short-duration bursts has been assumed.  $\eta$ can be
understood as an indicative of the terminal bulk Lorentz factor,
$\Gamma$.  The 
above  $\eta$ estimates assume, also for simplicity, that the  energy 
deposition rate by $\nu\overline{\nu}$ annihilation into \p pairs at
$t_{\rm sim}$ is both representative of the subsequent phases and steady
for {\it at least} one second (this is clearly an upper limit to the
total injected energy over the assumed period $\tau_{\rm inj}$). Due 
to the finite resolution of the simulations the densities along the rotation
axis are determined by particles located in the inner parts of the disk. 
We therefore expect to {\em overestimate} the densities
along the rotation axis. Given the above assumptions, higher 
numerical resolution would yield higher values for $\eta$.
This is supported by calculating these densities for two equivalent runs
of different numerical resolution (run A and B of RD). The densities in 
well-resolved case are lower by approximately one order of magnitude.\\
\begin{figure}
\centerline{\psfig{figure=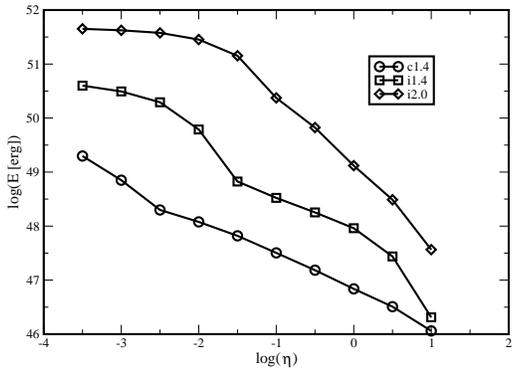,angle=-90,width=8cm}}
{\caption{Energy deposited via $\nu\overline{\nu} \rightarrow e^{+}e^{-}$ 
 as a function of $\eta$ for NS-NS merger models c1.4 (corotation, 2
$\times$ 1.4 $M_{\odot}$), i1.4 (no spins, 2 $\times$ 1.4
$M_{\odot}$),  i2.0 (no spins, 2 $\times$ 2.0 $M_{\odot}$).}
\label{fig2}}
\end{figure}
\begin{figure}
\centerline{\psfig{figure=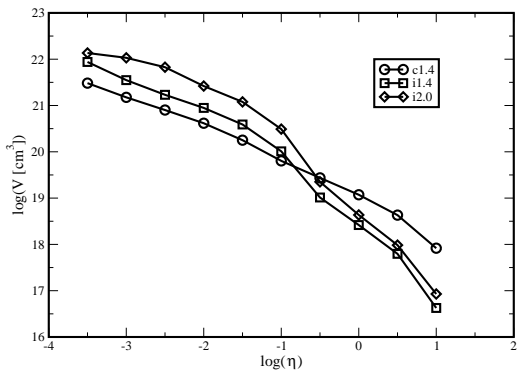,angle=-90,width=8cm}}
{\caption{Volume of the compact region containing material with a
given $\eta$ for NS-NS merger models c1.4,i1.4, i2.0.} 
\label{fig3}}
\end{figure}
The resulting constraints on the injected \p energy as a function of 
$\eta$ are illustrated in Figure 2. It shows the fast drop of the contained
energy as a function of the maximally reachable Lorentz factor. 
The copiously emitted electron neutrinos and
anti-neutrinos dominate the annihilation process by $\sim$ 95 \%,
their mean energies $\langle E \rangle$  are $\sim 10$ MeV for $\nu_e$
and $\sim 15$ MeV $\overline{ \nu_{e}}$. The highest
annihilation rates occur in the uninteresting (since extremely
baryon-rich) hot inner disk regions.  Even in the 
most optimistic case (i2.0) only  $\sim 10^{48}\;
(4\pi/\Omega_{\nu\overline{\nu}})$  erg  are found able to blow out
material near the axis with a Lorentz factor larger than 10 (indeed no  
SPH-particles are enclosed in this region, but we are limited in
our density calculation by the finite
resolution). A broad spread in the Lorentz factor will therefore be present in
the outflow -- close to the rotation axis $\Gamma$ may be high
(i.e. $\ge 10$); at large angles away 
from the axis an increasing degree of entrainment  will correspond to a
decrease in $\Gamma$ (see Fig. 1). 
Even if the outflow is not narrowly collimated,
some beaming is expected because the disk geometry (see Fig. 15 RD)
channels the energy  preferentially
along the rotation axis. But unless $\Omega_{\nu\overline{\nu}}$ is
$\le 10^{-1} -10^{-2}$ this may fail to satisfy the apparent
isotropized energies of $\sim 10^{51}$ erg implied for ``short-hard''
bursts at  a redshift $z=1$ (Panaitescu, Kumar \& Narayan 2001;
Lazzati, Ramirez-Ruiz \& Ghisellini 2001).

\subsection{Beaming and Confinement}

When a large amount of energy is suddenly released into a compact
region, an opaque ``fireball'' is created due to the prolific creation
of \p pairs (Cavallo \& Rees 1978; Goodman 1986;
\Pacz 1986; Shemi \& Piran 1990). A diagram similar to Figure 2 can be
drawn to illustrate  the volume that contains material
above a given $\eta$ (see Fig. 3). Consider a homogeneous fireball of
energy $E_{\nu\overline{\nu}}$, total mass $M_0$ initially confined
to a compact region, whose  volume is given by $V_{\rm
cav}$. Clearly, since the optical depth $\tau > 1$, the initial fireball 
will be an opaque sphere in thermal equilibrium, characterised by a
single temperature: 
\begin{equation}
T_{\nu\overline{\nu}}  \approx 10
E_{{\nu\overline{\nu}},48}^{1/4} V_{{\rm cav},18}^{-1/4}\;\; {\rm
MeV}, 
\end{equation}
where we adopt the convention $Q = 10^x\,Q_x$, using cgs units.
This radiative sphere expands and cools rapidly until the
energy of the photons degrades below the \p production threshold. When
some amount of baryonic matter is mixed with the fireball and the
radiation energy dominates the evolution (i.e. $\eta \gg 1$), the
fluid expands under its own pressure such that its Lorentz factor grows
initially linearly with radius, $\Gamma \propto r$, and reaches a final 
value of $\Gamma \approx \eta$ (Piran 1999).

If the flow is
assumed to be stationary and isentropic, then the fluid velocity will
increase as the external pressure of the surrounding medium
decreases. When the pressure has halved the
flow becomes transonic and the cross-sectional area is minimised. In
this way a directed de Laval nozzle can be established. Unlike
the static situation envisaged by Blandford \& Rees (1974), the
\p injected in this plasma do already have a  relativistic outward motion and
the flow is then unlikely to be recollimated in a fluid nozzle. Although it
will become free and supersonic (indeed relativistic, if  it  was so
initially) since the pressure drop at  $r_{\rm cav}$ is very steep. In
the simulations, the external pressure is found to vary with distance along the
rotation axis as $p_{\rm ext} \propto r^{\approx - 4}$.  If the
pressure drops faster than $r^{-2}$ onwards, sound waves will eventually be
unable to cross the jet, which will become overpressured with respect
to the external medium, and thenceforth expand out over an angle
$\Gamma^{-1}$. If the  free expansion starts just outside $r_{\rm
cav}$, where $\Gamma \sim 2$, then it will spread over a wide angle
(i.e. $\theta_{\nu\overline{\nu}} \sim 30^{o}$) and develop into a
roughly semi-spherical blast wave. The fireball 
will then accelerate until the entire energy is converted into kinetic
energy at $\eta r_{\rm cav}$.  If $\eta > \eta_b \approx 150
E_{{\nu\overline{\nu}},48}^{1/3} r_{{\rm cav},6}^{-2/3}$, the fireball
continues to be radiation dominated when it becomes optically thin and
most of the energy escapes as photons. An observer will detect them
with a characteristic thermal peak frequency of  $T_{\rm obs} \approx
T_{\nu\overline{\nu}}$ (Goodman 1986), which is likely to be outside the
BATSE [20-600]~keV 
spectral window (see equation 1). Alternatively, if $\eta < \eta_b$,
the fireball becomes matter dominated, and most of the initial energy
is converted into bulk kinetic energy with $\Gamma \approx \eta <
\eta_b$. The inertia of the swept-up external matter will then
decelerate these ejecta  significantly by the time it reaches $r_\Gamma
\approx 10^{16} E_{{\nu\overline{\nu}},48}^{1/3} \eta_{1}^{-2/3}
\theta_{{\nu\overline{\nu}},1}^{-2/3} n_{{\rm ism},0}^{-1/3}$ cm
(M\'esz\'aros \& Rees 1997). For an approximately smooth distribution
of external matter, the bulk Lorentz factor of the fireball decreases
as an inverse power of the time ($\propto t^{-3/8}$). As a
consequence, the spectrum softens in time as the synchrotron peak
corresponding to the minimum Lorentz factor and magnetic field
decreases,  leading to the
possibility of multi-wavelength afterglow radiation. 

\subsection{Accompanying winds}

The neutrinos that are emitted from the inner shock-heated regions of the
torus will deposit part of their energy in the outer parts of the thick disk
that has formed around the central object (see Fig. 15 of RD). 
The possibility of such a wind from a NS coalescence has 
already been realized  by Ruffert et al. (1997). 
Using the typical numbers from our simulations, we find that the neutrinos will
drive a mass outflow  at a rate (Qian \& Woosley 1996):  
\begin{equation}
\dot{M}\approx 2 \cdot 10^{-2} {\rm M}_{\odot}/{\rm s}\;
L_{{\overline{\nu}},53}^{5/3} 
\left(\frac{\epsilon_{\overline{\nu}_e}}{15 {\rm MeV}}\right)^{\frac{10}{3}}  
\left(\frac{R_{\rm em}}{80 {\rm km}}\right)^{\frac{5}{3}}
\left(\frac{2.5 M_{\odot}}{M_{\rm em}} \right)^2 \nonumber,
\end{equation}
where $L_{\overline{\nu}_e}$ is the luminosity in
$\overline{\nu}_e$-neutrinos, 
$\epsilon_{\overline{\nu}_e}= \langle
\frac{E^2_{\overline{\nu}_e}}{E_{\overline{\nu}_e}} \rangle$, $R_{\rm
em}$ and  $M_{\rm em}$ are the radius of and the mass inside the neutrino 
emitting region. This estimate is   
consistent with those of other authors (Duncan, Shapiro \& Wassermann 
1986, Woosley 1993). This wind accelerates the blown-off material to its 
asymptotic velocities of $\sim 10^9$ cm s$^{-1}$. A wind of this strength
represents a possible danger for the emergence of a highly relativistic jet.
The jet can only accelerate to high Lorentz factors 
if the wind  is kept out of the funnel region along the rotation axis via
centrifugal forces. However, from the bulged geometry (see Fig. 15 RD) and the
assumption that the neutrinos  will be emitted preferentially in the direction 
$\hat{n}= -\nabla \rho /|-\nabla \rho|$, we conclude that most of the wind 
will be still directed away from the jet. 

\section{Discussion}
 
\subsection{$\nu\overline{\nu}$ annihilation as a viable energy source?}

We now want to address the viability of NS-NS binary coalescences as 
central engines of GRBs. The main form of energy release from the disk
we considered is neutrino emission, which produces relativistic jets
but fails to produce very  energetic bursts. 
Due to our post-processing approach the interaction between the jet 
and the matter evolution/neutrino emission is neglected. Not accounting
for the evolution of the jet during the energy injection period leads
to an overestimation of the energy density in the jet. But even 
under the most favourable conditions (maximum baryon free volume allowed by the
SPH-particle distribution, neglecting a neutrino driven 
wind that might pollute the funnel region above the poles, see \S 3, and
ignoring the jet evolution) 
we find typical energies of only $\sim  10^{48}$ erg, where we have assumed a
steady injection  lasting for as long as 1 s. Unless the \p plasma was
collimated into a solid angle $\Omega_{\nu\overline{\nu}} \le
10^{-2}$ (which is unlikely if the plasma is solely confined by gas
pressure; but see Levinson \& Eichler 2000 for an alternative
mechanism for hydrodynamic collimation), $\nu\overline{\nu}$
annihilation could only be 
responsible for a rather weak energy release (which is also likely to be
very short, almost impulsive; see Lee \& Ramirez-Ruiz 2002) and may 
therefore fail to produce isotropized energies of $\sim 10^{51}$ ergs that
are estimated for ``short-hard'' bursts if they are at $z \sim
1$. Even if  this is the case, the 
\p deposited by  $\nu\overline{\nu}$ annihilation  may be responsible
for either precursor emission (thermal and non-thermal) and/or for
creating a cavity into which a magnetised jet can subsequently expand (see
below). In this latter case, the afterglow, at least over a range of
directions, would not arise until the late magnetised ejecta hits the
wall of the cavity. One would naturally expect  a wide variety of afterglow
behaviours arising from the impact of (possibly anisotropic) ejecta
on an irregularly shaped cavity.\\ 
It is, of course also  possible to produce more energetic bursts via 
$\nu\overline{\nu}$ annihilation if one is able to increase the
temperatures in the  appropriate density regimes (i.e. $\sim 10^{11}$ to
$10^{13}$ gcm$^{-3}$).  One such possibility would be that the ``real''
equation of state is much softer 
than the one used here, leading to higher temperatures and
hence more energetic  neutrino emission. The neutrino
luminosities are sensitive functions of the temperature of the emitting
region, the $\beta$-reactions are $\propto T^6$ and the plasmon decay
is $\propto T^9$. Higher temperatures would translate into higher 
neutrino energies $E_{\nu}$, which in turn render the debris more opaque 
to the neutrinos (the cross-sections scaling $\propto E_{\nu}^2$). We
still, however,  expect higher luminosities and hence higher annihilation
energies if the temperature is increased. An alternative possibility may
arise from general relativistic effects, which are 
not included in the current 3D hydrodynamics simulation. 
The tendency of general 
relativistic effects would be to compactify the matter distribution,
and so higher temperatures would be naturally expected. 
However, in this case the annihilation
process would be additionally  complicated by effects such as bending of
neutrino trajectories and redshifts of the neutrino energies.
These effects have been found to partly cancel out one another so that
they do not alter the results substantially. For a further discussion of
this point we refer to the literature (Salmonson \& Wilson 2001, 
Asano \& Fukuyama 2000, Jaroszynski 1993).
A third possibility could arise from a very high disk viscosity. In our
current simulation no explicit, physical viscosity has been introduced. We 
measured an equivalent $\alpha$-viscosity of the order of $\alpha_{\rm
disk} \approx 4 \cdot 10^{-3}$ (lower 
values are found in the better resolved central object; see
RD). The reason for this  low value is a largely improved artificial
viscosity scheme and the high  resolution. Future investigations
including explicit, physical viscosity should clarify this point. 
The raised possibilities should certainly be kept in mind, but at the
present stage they are a matter of speculation.

\subsection{Delayed BH formation and MHD jets}

An alternative way to tap the debris  energy is via magnetic fields
threading the torus. Even before a BH forms, the merging
system might lead to winding up the fields and dissipation before the
merger.  While the above mechanism can tap the rotational energy
available in the debris torus, a BH formed from the coalescence is
guaranteed to be rapidly rotating, and being more massive, could
contain a larger energy reservoir than the debris itself. A magnetic
configuration capable of powering bursts requires fields of 
a few times $10^{15}$ G. A weaker field would extract inadequate
power; however, it only takes of order of a second for simple winding
to amplify a  $10^{12}$ G field to $10^{15}$ G (Klu\'{z}niak \&
Ruderman 1998). If magnetic
fields of comparable strength than those anchored in the torus thread
the BH, its rotational energy offers an extra (an possibly even greater) source
of energy than in principle can be extracted via the B-Z mechanism
(Blandford \& Znajek 1977). For a maximally rotating BH, this is
0.29 $M_{BH} c^2$ erg, multiplied by some efficiency factor. Even
allowing for low total efficiency ($\sim$ 20\%), a jet powered by the
B-Z effect would not require any beaming to produce the equivalent of
an isotropic energy of $10^{53.5}$ erg. The entrained baryonic mass
would then only need to be below  $10^{-4} M_\odot$ (which is clearly
the case along the rotation axis of the merger remnant) to allow high
relativistic expansion speeds.  \\
The masses of the central objects of the merger
remnant are $\sim$ 2.5 M$_{\odot}$ and therefore above the value of
$\sim 2.2$ M$_{\odot}$ that is typically found for cold, non-rotating
NS  in  beta-equilibrium (see e.g. Akmal et al. 1998). 
Due to the poorly known physics under these extreme conditions (strong field  
gravity, magnetic fields, 'exotic' matter in the NS cores) the time 
scale on which the central object is going to collapse is, however, uncertain.
For the generic NS merger case, i1.4, there are no signs
of a collapse -- at the end of the simulation, the maximum density of
the merger remnant has not even reached the central density of a cold,
non-rotating single 1.4 M$_{\odot}$ NS. This is mainly due to the stabilising
effect of differential rotation. For a discussion of further stabilising 
effects (some of which are not included in the simulation) we refer to
RD. The merger remnant would lose angular momentum on a magnetic dipole 
radiation time scale  
\begin{equation}
\tau_c \sim 10^2 \; {\rm s} \left(\frac{M_{\rm CO}}{2.5 \;M_{\odot}} \right) 
\left(\frac{10^{16} \;{\rm G}}{B} \right)^2 \left(\frac{15 \; {\rm km}}
{R_{CO}}\right)^4
\left(\frac{3000}{s \cdot\omega} \right)^2,
\end{equation}
which is substantially shorter than the viscous timescale, $10^9$ s 
(Shapiro 2000).
Due to the dependence of this timescale on the input quantities, we expect
a large range of possible delays between the merger and the subsequent
BH formation. This will be mainly determined by the details of the
coalescence, unless exotic matter appears already for very low masses,
therefore  leading to an immediate collapse. 
Note that with reasonable assumptions even a time scale of weeks is
still possible. More prominent time delays could also be expected for
mergers of unequal mass NS, or NS with other compact companions. Such
events might also lead to repeating episodes of accretion, or to the
eventual explosion of the NS which has dropped below the critical
mass, all of which would provide a longer lasting energy output. A GRB
produced by MHD coupling is likely to go off inside a cavity inflated by the
$\nu\overline{\nu}$ annihilation burst (which may not produce a
standard GRB; see \S 3). Such cavities 
can be as large as fractions of a parsec, giving rise to a
magnetic deceleration shock months after the first  $\nu\overline{\nu}$
annihilation burst. \\ 
Much progress has been made in understanding how gamma-rays can arise
in fireballs produced by brief events depositing large amounts of
energy in a small region, and in deriving the generic properties of
the corresponding afterglows. Nonetheless, there still remain a
number of mysteries, especially concerning the progenitors of
short-duration bursts. As argued above, a wide diversity of behaviours
may  be the rule, rather than the exception. And if the class of
models that we have advocated here turns out to be irrelevant, the
trigger of these events has to be even more
remarkable and exciting.

\section*{Acknowledgements}
We are grateful to M. J. Rees for numerous insightful comments and
suggestions. It is a pleasure to thank W. H. Lee and J. Salmonson for
discussions and the Leicester supercomputer team S.
Poulton, C. Rudge and R. West for their excellent support. 
The computations reported here were performed using both the UK
Astrophysical Fluids Facility (UKAFF) and the University of Leicester 
Mathematical Modelling Centre's supercomputer. This work was supported by
PPARC, CONACyT, SEP and the ORS foundation.

\end{document}